\documentclass[11pt]{article}

\usepackage{graphicx}
\usepackage{dcolumn}
\usepackage{bm}

\usepackage[utf8]{inputenc}
\usepackage[T1]{fontenc}
\usepackage{mathptmx}
\usepackage{etoolbox}
\usepackage{amsmath,amssymb}
\usepackage{color}

\DeclareMathOperator{\erfc}{erfc}
\newcommand{\dfnd}{\stackrel{\mathrm{def}}{=}}
\newcommand{\dx}{
  \left| \Delta \vec{x}_{\bot} \right|^2
}

\newcommand{\diff}{\,\mathrm{d}}

\begin{document}


\title{Numerical study of the Transverse Diffusion coefficient for a one
  component model of plasma}
\author{Lorenzo Valvo\thanks{
  Department of Mathematics, Universit\`a degli Studi di Milano,
  Via Saldini 50, 20133 Milano, Italy. Email: \texttt{lorenzovalvo@gmx.com} }
  \and
  Andrea Carati\thanks{%
  Department of Mathematics, Universit\`a degli Studi di Milano,
  Via Saldini 50, 20133 Milano, Italy. E-mail: \texttt{carati@mat.unimi.it}
  }%
}

\date{\today}

\maketitle

\begin{abstract}
  In this paper we discuss the results of some Molecular Dynamics
  simulations of a magnetized One Component Plasma, targeted to
  estimate the diffusion coefficient $D_{\perp}$ in the plane
  orthogonal to the magnetic field lines.  We find that there
  exists a threshold with
  respect to the magnetic field strength $|\vec B|$: for weak magnetic
  field the diffusion coefficients scales as $1/|\vec B|^2$, while a
  slower decay appears at high field strength.  The relation of this
  transition with the different mixing properties of the microscopic
  dynamics is investigated by looking at the behavior of
  the velocity auto correlation.
\end{abstract}


\noindent
\textbf{
  The diffusion process is well understood for stochastic
  motions (see Ref.~\cite{balescu_aspects_2005}), that are supposed to
  mimic the behaviour of a chaotic dynamical system. Many questions
  are instead left open in the study of the diffusive properties of a
  system which is in a partially ordered state (see for example
  Ref.~\cite{ZASLAVSKY2002461}).
    A central issue, as regards magnetized plasma confinement, is
  the diffusion of charged particles in the direction perpendicular to
  the magnetic field lines. A widely accepted law, predicting that the
  transversal diffusion $D_{\perp}$ coefficient is proportional to the
  inverse of the square of the magnetic field strength $|\vec B|$, was
  proposed more than 50 years ago( see
  Ref.~\cite{longmire_diffusion_1956}). Being based on kinetic theory,
  this law is expected to hold whenever the microscopic dynamics is
  chaotic.
  However, as the magnetic field $|\vec B|$ is increased, a
  partially ordered state seems to set in (see
  Refs.~\cite{carati_chaoticity_2014,carati_transition_2012}), at
  least for a pure electron plasma. Our purpose was to investigate the
  consequences (if any) of this transition on the diffusion process.
  So we have performed Molecular Dynamics simulations of a
  magnetized one component plasma, that is a set of mutually
  interacting electrons subject to a constant external magnetic
  field. We estimate the diffusion coefficient $D_{\perp}$ in the
  plane orthogonal to the field, for different values of the magnetic
  field strength $|\vec B|$.  We find that the kinetic law holds for
  low $|\vec B|$, when the microscopic dynamics is chaotic. But as the
  magnetic field grows the diffusion coefficient seems to saturate to
  a plateau, while the microscopic state turns to a partially ordered
  one.
}

\section{\label{sec:intro} Introduction}

In the years Sixties there was a great exchange among
research groups of plasma physics and of dynamical systems,
as both were interested in the study of the $1\tfrac{1}{2}$
Hamiltonian system that can represent the magnetic field lines;
see Ref.~\cite{escande_contributions_2016}
for a broad historical overview. In more recent years, another
point of connection have become the study of diffusion process
in the phase space, see Ref.~\cite{balescu_aspects_2005}.
It has been shown that if the dynamics is not fully chaotic, then the process
of diffusion in the phase space can be ``not normal'', i.e., the
mean square displacement doesn't necessarily grow linearly in time:
the process is called super diffusive if the growth is faster than
linear, or sub diffusive otherwise; for a review
see~\cite{ZASLAVSKY2002461}. In both cases there is no
widespread agreement of the correct definition of the diffusion
coefficient to be adopted.
\begin{figure}
  \centering
  \includegraphics[width=\textwidth]{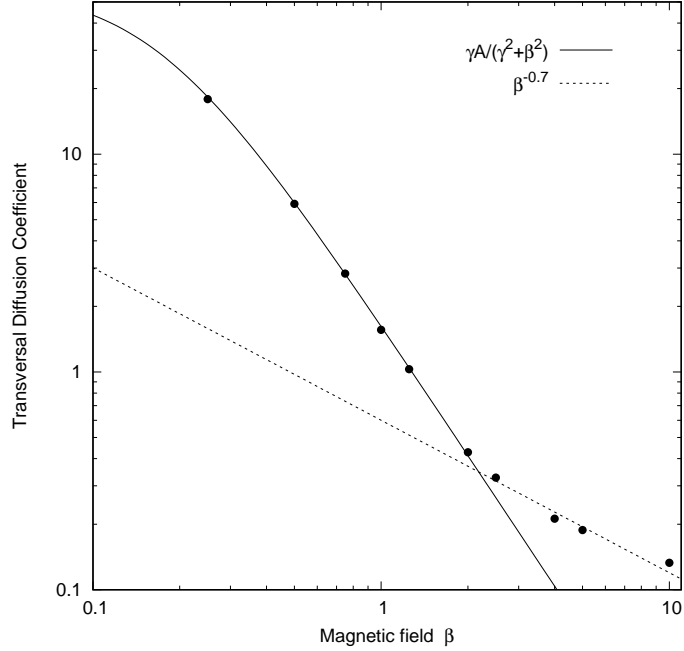}
  \caption{ Diffusion coefficient $D_\perp$ perpendicular to the
    magnetic field versus $\beta$, computed by MD simulations. Circles
    are the numerical results, while the solid line is the plot of the
    function $D_\perp = \frac {\gamma A}{\gamma^2 + \beta^{2}}$, with
    $\gamma=0.168$ and $A=9.9$.  Such values are obtained from the
    auto correlation of the electron velocities, as explained in the
    text. The broken line is the plot of the function $D_\perp\simeq
    \beta^{-0.7}$,  corresponding to the law found in
      paper \cite{zimbardo}.
   }\label{fig:graph_D_vs_beta}
\end{figure}

Now, as it will be shown below, the equations of motion for a
single electron in a one component plasma subject to an external
constant magnetic field $\vec B$ can be recasted in a dimensionless
form, in which the only parameter appearing is the ratio
$\beta=\sqrt{4\pi}\omega_c/\omega_p$ among the cyclotron frequency
$\omega_c=eB/mc$ and the plasma frequency $\omega_p=\sqrt{4\pi ne^2/m}$
(in the c.g.s. system); here $n$ is the particle density and $e$ the
electronic charge. The parameter $\beta$ measures the relative
strength the magnetic Lorentz force acting on a single electron, with
respect to the electrostatic force due to all other electrons. In the
limit of $\beta\to +\infty$ the equations of motions decouple and the
system reduces (formally) to a set of independent electrons in a
constant magnetic field, i.e., to an integrable system.

So, it can be conjectured that for high magnetic field strength, the
dynamics will not be fully chaotic, (see papers
\cite{carati_chaoticity_2014} and \cite{carati_transition_2012}), and
that the diffusion process in phase space may be ``anomalous''.
Actually, it is impossible for us to study numerically such process,
and we limit ourselves to study the diffusion of the electrons in the
physical space. To this end we study the diffusion coefficient,
defined as usual (see for example Ref.~\cite{allen}) by
\begin{equation}
  \label{eq:diffusionCoefficientDef}
  D_\perp = \lim_{t\rightarrow\infty} \frac{\langle\dx\rangle}{4t}
\end{equation}
being $\dx$ the mean displacement, in the plane perpendicular to the
magnetic field, of the electrons from their initial positions, while
$\langle \cdot \rangle$ should be the phase average. Instead in this
paper, following a common attitude, averages will be always taken as
time averages along the orbits. We have not investigated the relations
between the two averages.

As regards physical applications, small values of
$D_\perp$ are important for the purpose of plasma confinement in
fusion devices.  This is another reason to investigate in which regime
the diffusion coefficient is small.

Another dimensionless parameter which characterize the state of a
plasma is the coupling parameter $\Gamma=e^2/(ak_B T)$, where $a$ is
the inter particle spacing (related to the particle density $n$ by the
relation $a=n^{1/3}$), $T$ the plasma temperature and $k_B$ as usual
the Boltzmann constant. So defined, $\Gamma$ is the ratio among the
mean coulomb energy of a couple of nearest particles, and the mean
kinetic energy. The weak coupling regime is then defined by $\Gamma
<1$ and the strong coupling regime by $\Gamma\geq 1$. Up to now,
because of the reasons explained in the following, Molecular Dynamics
(MD) simulations have been performed mainly in the strongly coupled
case, while the weakly coupled regime have been addressed mostly by
kinetic theory.

In literature it is possible to find different estimates for the
diffusion coefficient. The oldest one (see
Ref.~\cite{longmire_diffusion_1956}) predicting the scaling law
$D_\perp\propto \beta^{-2}$, is obtained in the frame of the kinetic
theory, in the weak coupling regime.  However, other different
theories have been proposed in the years, each giving a different law
for the dependence of the diffusion coefficient $D_\perp$ on the
parameter $\beta$. A few of them are summarized in Table~1 on page
135003-2 of Ref.~\cite{ott_diffusion_2011}; another one is percolation
theory (see Refs. \cite{isichenko,isichenko2,zimbardo}), which
predicts a scaling like $\beta^{-0.7}$. This law is in the closest
agreement with our results. It was brought to our attention by an
anonymous referee that we warmly thank.

But none of the proposed theories is based on the loss of chaoticity
in the Newtonian microscopic dynamics. Also the numerical works found
in literature fail to address this point. In fact, up to now, the
behaviour of the diffusion coefficient have been investigated by MD
mostly in the unmagnetized case, see for instance
Refs.~\cite{daligault_diffusion_2012,saumon_diffusion_2014,caplan_precise_2021}.
The magnetized case was studied in Ref.~\cite{ott_diffusion_2011}, but
only in the strong coupling regime: at the smallest value $\Gamma=1$ a
transition at $\beta\approx 1$ in the behavior of $D_\perp$ was
observed, but the origin of such a transition was not discussed.  A
similar transition was observed also in two more recent works
\cite{baalrud_transport_2017} and \cite{vidal_extended_2021}, where
the diffusion coefficient was studied numerically for $\Gamma$ down to
$0.1$. However, those works were based on a so-called
Particle-Particle Particle-Mesh ($P^3M$) code, which is a sort of
hybrid between a kinetic and a MD code. We think that such method is
not suited to study the chaoticity of the microscopic dynamics. More
on the connections between our results and the cited paper will be
said in the conclusions.

So in this paper we perform a fully MD simulation for different values
of $\beta$ for a system of $N=4096$ electrons, for a fixed value of
$\Gamma=0.1$, which is the smallest value we were able to manage. The
aim is to verify for what value of $\beta$ the transition in the
behavior of $D_\perp$ occurs, and to observe the 
chaoticity of the dynamics, by computing the time auto correlation of
the transverse particle velocity.

In Section~\ref{sec:model} we describe the model and the numeric
algorithm used, in Section~\ref{sec:back} the numerical results are
reported while in Section~\ref{sec:conclusions} the conclusions
follow.

\section{\label{sec:model}The model and the numerical scheme}

The system we are considering is called in the literature a one
component plasma and it consists of a number $N$ of electrons in a
cubic box of side $L$ with periodic boundary conditions, the electrons
being subject to mutual Coulomb interactions, and to an external
constant magnetic field $\vec{B} = B\vec{e}_z$ ($\vec{e}_z$ is the
unit vector directed along the $z$ axis).  The density is then defined
by $n=N/L^3$. This is considered a model of a plasma as the positive
ions are supposed to constitute a uniform neutralizing background.

If $t$ denotes time and $\vec{x}_i$ the position of the $i$-th electron
(with $i=1,\dots,N$), we define dimensionless variables
\begin{equation}
  \vec{y}_i = a^{-1}\vec{x}_i,\quad \tau = \omega_c t \ ,
  \label{units}
\end{equation}
by rescaling distances with the inter particle spacing $a=n^{-1/3}$ and
time with the cyclotron frequency $\omega_c$. Using such variables,
the equations of motion for the $i$-th electron read
\begin{equation}
  \label{newtonEquation}
  \frac{d^2 {\vec{y}}_i}{d\tau^2} = \vec{e}_z \times \frac{{d\vec{y}}_i}{d\tau}
    +\frac{1}{\beta^2} \vec{E}(\vec{y}_i)
\end{equation}
where $\vec{E}(\vec{y}_i)$ is the electric field acting on $i$-th
electron due to all other charges. The electric field of a periodic
system of charges can be computed via the Ewald formula
(see for example Ref.~\cite{gibbon_long-range_2002}),
\begin{equation}
  \begin{split}
    &\vec{E}(\vec{y}_i) = \\
      & \sum_{\vec{l}} \sum_{j=1}^N
        \frac{ \vec{y}_{ij\vec{l}} }{| \vec{y}_{ij\vec{l}} |^3}
	\left[ \erfc({\alpha} | \vec{y}_{ij\vec{l}} | )
        + \frac{2 \alpha | \vec{y}_{ij\vec{l}} | }{ \sqrt{\pi} }
        \exp(-{\alpha}^2{ | \vec{y}_{ij\vec{l}} |}^2) \right] \\
      &  + \frac{4 \pi}{N} \sum_{\vec{q} \neq 0} \sum_{j=1}^N
        \frac{\vec{q}}{q^2} e^{-q^2\!/4 {\alpha}^2}
        \sin( \vec{q} \cdot \vec{y}_{ji}),
	\qquad \alpha = \sqrt{\pi}N^{-1/6} 
	\label{ewaldField} \\
  \end{split}
\end{equation}
where $\vec{l}$ is a triplet of integers denoting the position of an
image cell, while $\vec{q}$ is a vector in the reciprocal lattice,
i.e. is defined by $\vec{q}=2\pi\vec{n}/L$ with $\vec{n}$ an integer
vector, and finally we have defined $\vec{y}_{ij\vec{l}} = \vec{y}_i -
\vec{y}_j + \vec{l}\sqrt[3]N$. The two series are truncated as to
assures a relative error below $10^{-7}$, which is smaller than the
relative error of the energy conservation in a single numerical step.

Of the two dimensionless parameters of the problem, only $\beta$
appears into the equations of motion. The parameter $\Gamma$ enters
through the choice of the initial data: in fact, while the positions
are extracted from a uniform distribution, the velocities are taken
from a Maxwell distribution, and the temperature is uniquely
determined by $\Gamma$.  With this choice, at the beginning of each
simulation, the system is out of equilibrium: so there is a drift of
the kinetic energy, and the system reaches a different, random,
temperature.  In order to fix the temperature to the desired value, we
operate in this way: after extracting the initial values, we let the
system evolve until equilibrium is reached, i.e. until the kinetic
energy appears to stabilize. We then generate new velocities again
with a Maxwell distribution at temperature $T$, and repeat the process
until the kinetic energy appears to be constant and close to the
chosen value.

The equations \eqref{newtonEquation} were integrated using a
symplectic splitting algorithm (see for example Ref.~\cite{Hairer}).
The total Hamiltonian
\begin{equation}
  H=\frac 12 \sum_k \Big( \vec p_k +\frac 12 \vec e_z\times\vec y_k
  \Big)^2 + \frac 1{\beta^2}V(\vec y_1,\ldots,\vec y_N) \ ,
\end{equation}
where $V(\vec y_1,\ldots,\vec y_N)$ is the Coulomb potential of the
electrons computed according the Ewald prescription,
was split as $H=H_1+H_2$ where
\begin{equation}
  H_1\dfnd \frac 12 \sum_k \Big( \vec p_k +\frac 12 \vec
  e_z \times\vec y_k \Big)^2 \ ,\quad H_2\dfnd V(\vec y_1,\ldots,\vec y_N)
  \ .
\end{equation}

\begin{figure}
  \centering \includegraphics[width=\textwidth]{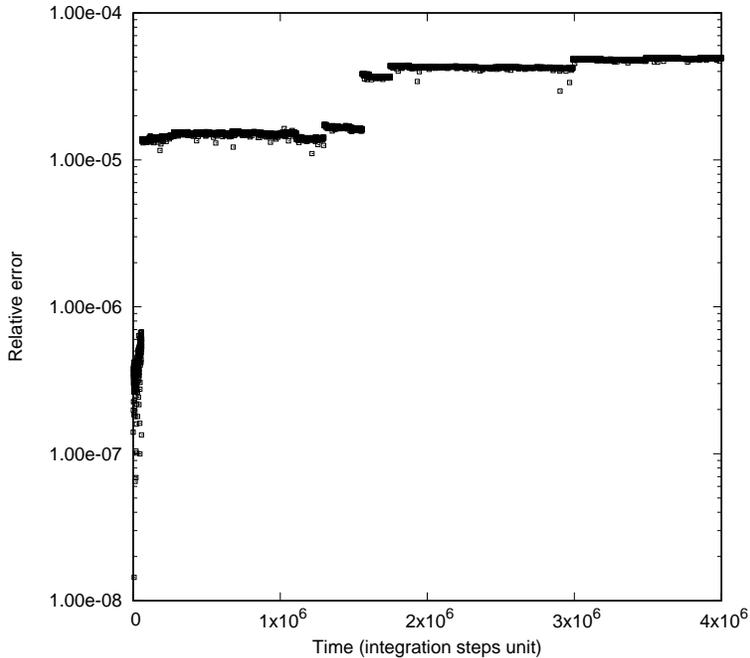}
  \caption{ Relative error of the energy conservation as a function of
    the number of integration steps.  Data refers to a case with
    $\beta=1$ and $N=4096$ particles. }\label{fig:enrgcons}
\end{figure}

Now, denoting by $\Phi^t_1$ the flow determined in the phase space by
the Hamiltonian $H_1$ and by $\Phi^t_2$ the one due to the Hamiltonian
$H_2$, the integration algorithm is obtained simply by composition
$\Phi\dfnd\Phi^{\delta t/2}_1\circ\Phi^{\delta t}_2\circ\Phi^{\delta
  t/2}_1$, where $\delta t$ is the time step. Such time step was
chosen so that the energy conservation was better then a part over
$10^{4}$ in all the simulations.  In particular we choose $\delta
t=2\pi C(\beta)\,10^{-4}$, where the factor $C(\beta)$ is taken equal
to $\beta$ for $\beta<1$, and costant equal to 1 for larger $\beta$.
In figura~\ref{fig:enrgcons} we report the relative error of the
energy conservation as a function of the number of integration step,
for a tipical run.  The number of step used was $4\,10^6$ for
$\beta\le 2.5$, four times this number for $\beta$ larger.

In MD simulations the choice of the number $N$ of particles is always
an issue, all the more in the context of a weakly coupled Plasma. For
a very rough estimation, we made the following considerations: the
fundamental cell of the simulation should have a side larger than the
Debye length $\lambda_D$, which in our units reads $\lambda_D =
\sqrt{1/\Gamma}$.  The first of \eqref{units} implies that $L =
N^{1/3}$, so that the requirement $L > \lambda_D$ in our units becomes
$N > \Gamma^{-3/2}$. As the Coulomb force is long range, the
computational cost scales as a power of $N$.  With a clever use of the
Ewald summation formula, see Ref.~\cite{Perram}, the
computational cost scales (asymptotically) as\footnote{It is possible
  to conceive algorithms with an even slower asymptotic growth, but
  for the order of magnitude of $N$ that we are considering, the Ewald
  summation formula works best.} $N^{3/2}$. So we cannot afford to
choose a value much bigger than $N=4096$. In any case, for $\Gamma=
0.1$ our constraint reads $N > 32$ and so with our choice of $N=4096$
we have $L \simeq 5 \lambda_D $.

Another interesting length scale which appears in the problem is the
Larmor radius $r_l$, i.e., the gyration radius of the electrons
determined by the magnetic Lorentz force due to the external magnetic
field $B$.  A simple computation shows that one has
$r_l/L=\sqrt{2/\Gamma}/\beta N^{1/3}$. For the smallest value of
$\beta$ used in our computations, i.e. $\beta=0.25$, the Larmor radius
is slightly larger than the side of the simulation cell, because one
gets $r_l\simeq 1.1 L$. Instead $r_l$ turns out to be well below the
Debye length, for the largest value of $\beta=10$.

\section{\label{sec:back}Results}

We recall that the transverse diffusion coefficient is defined by
\eqref{eq:diffusionCoefficientDef}, where
\begin{equation}
  \dx \dfnd \frac {\sum_k \Big(|x_k(t) - x_k(t_0)|^2 + |y_k(t) -
    y_k(t_0)|^2\Big)}N
\end{equation}
is the mean particles displacement in the plane orthogonal to the
magnetic field $\vec B$ (we recall that the latter
was taken directed as the $z$--axis) and the brackets
denote the time average along the
orbit. To estimate this quantity we proceed as follows.

Let $\delta t$ be the integration step and $M_{tot}$ the total number
of integration steps. Let also $j$ be an integer smaller than a fixed
fraction $M'$ of $M_{tot}$ (we take one sixteenth), then the time
averages of $\langle\dx\rangle$ at time $t_j=j\delta t$ were computed
as
\begin{equation}\label{eq:meandiffcoef}
  \langle\dx\rangle(t_j) = \frac 1M \sum_{h=1}^M \left( \frac 1N
  \sum_{k=1}^N \Big(|x_k(t_{h+j}) - x_k(t_{h})|^2 + |y_k(t_{h+j}) -
  y_k(t_h)|^2\Big)\right) \ ,
\end{equation}
where $M=M_{tot}-M'$.  After plotting $\langle\dx\rangle$ as a
function of time, we fit the tail with a straight line whose angular
coefficient (divided by four) is an estimate of the diffusion
coefficient. The values computed in such way can be found in
Table~\ref{table:uno} (second entry).
\begin{table}[t]
  \caption{\label{table:uno} Values of the diffusion coefficient
    $D_\perp$, toghether with the values of the parameter $A$,
    $\omega$ and $\gamma$ appearing in the fit of velocity auto
    correlation.}
  \begin{center}
    \begin{tabular}{cccccc}
      \hline
      $\beta$ & ~$D_{\perp}$$~^{\mathrm{a}}$ & ~$D_{\perp}$$~^{\mathrm{b}}$  & $A$ & $\omega$ & $\gamma$   \\
      \hline
      0.25     & 18.1     &  17.9    & 9.90 & 0.253 & 0.169            \\
      0.50     & 5.61     &  5.91    & 9.85 & 0.506 & 0.169            \\
      0.75     & 2.68     &  2.83    & 9.80 & 0.754 & 0.167            \\
      1.00     & 1.56     &  1.63    & 10.1 & 1.01  & 0.164            \\
      1.25     & 1.03     &  1.07    & 9.95 & 1.258 & 0.166            \\
      2.00     & 0.428    &  0.434   & 9.90 & 2.022 & 0.165            \\
      2.50     & 0.327    &  0.342   & 9.75 & 2.518 & 0.171            \\
      4.00     & 0.212    &  0.200   & 10.0 & 4.046 & 0.168            \\
      5.00     & 0.188    & 0.184    & 10.35 & 5.035 & 0.164            \\
      ~~10.0~~ & ~0.133~~ & ~0.146~~ & 10.30 & ~~10.06~~ & ~~0.157~~    \\
      \hline
    \end{tabular}
  \end{center}
      {\small $~^{\mathrm{a}}$ computed from the linear fit of the tail of
        function (\ref{eq:meandiffcoef}).}
      
      {\small $~^{\mathrm{b}}$ computed from the velocity correlation
        by formula (\ref{eq:diffusione}).}
\end{table}

\begin{figure}
  \centering
  \includegraphics[width=\textwidth]{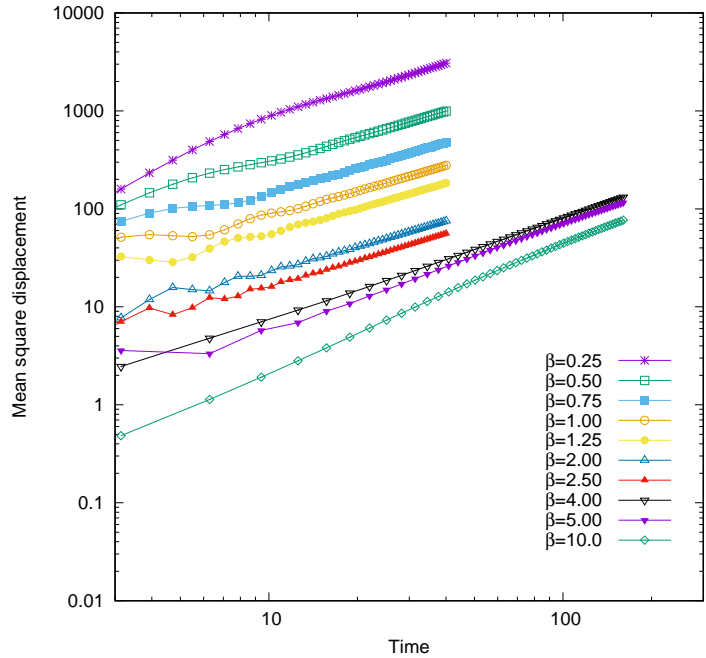}
  \caption{
    The average square displacement as a function of time. It is
    interesting to observe the initial non diffusive behaviour, which
    becomes more evident as $\beta$ is increased.
  }\label{fig:oscillations}
\end{figure}

The whole set of our plots (in log--log scale) can be found in
figure~\ref{fig:oscillations}; as usual, they display a diffusive
(linear) behaviour only after a certain time (the so called
``ballistic jump''). So we also restricted the sets of points
to the latter time window to exclude small times.

The results of our computations are summarized in
figure~\ref{fig:graph_D_vs_beta}. There, in logarithmic scale, the
numerically computed values of the coefficient $D_{\perp}$ are
reported (full circles) as a function of $\beta$.  All the simulations
were performed at the same value of $\Gamma=0.1$. The solid line is
the plot of the function $D_\perp = \frac {\gamma A}{\gamma^2 +
  \beta^{2}}$, with $\gamma=0.168$ and $A=9.9$. It can be checked that
this function agrees very well with the numerical results for
$\beta<2$. We remark that, for small values of $\gamma$, this function
decrees as $\beta^{-2}$, i.e., for $\beta<2$, the computed values of
$D_{\perp}$ agrees with the prediction of the kinetic theory.  But by
further increasing the magnetic field above a value about $2$, the
plot shows a ``knee'': the diffusion coefficient appears to obey a
different law. These results are in agreement with figure~5(a) of
Ref.~\cite{vidal_extended_2021}.  Notice that for such value of
$\beta$, the Larmor radius $r_l$ is slightly smaller than Debye
length.
\begin{figure}
  \centering
  \includegraphics[width=\textwidth]{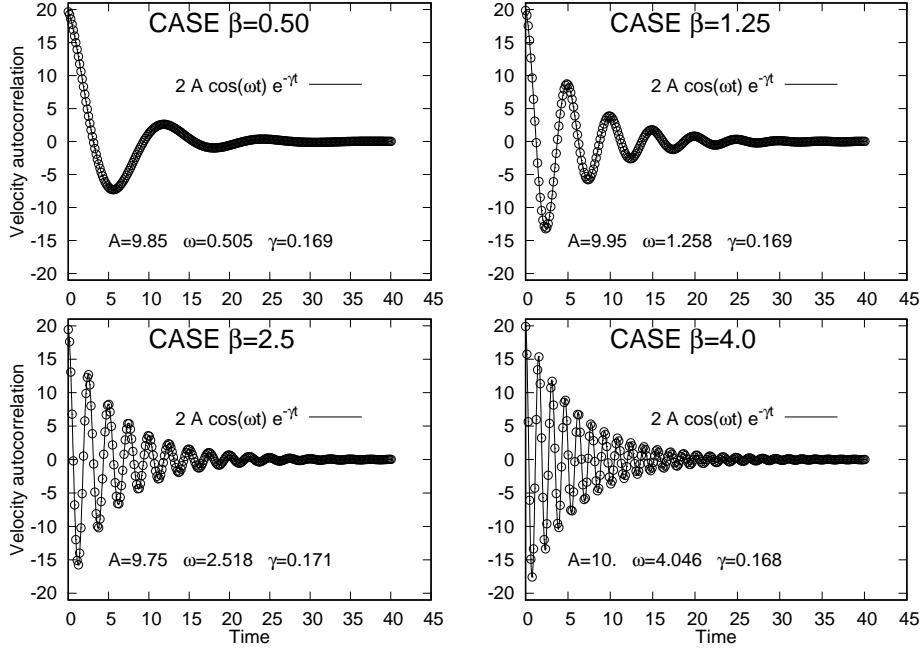}
  \caption{         Velocity          auto correlation         $\langle
    \vec{v}_{\perp}(t)\cdot\vec{v}_{\perp}(0)\rangle$,  as a  function
    of   time,  for   different   values  $\beta=0.5$,   $\beta=1.25$,
    $\beta=2.5$  and  $\beta=5$,  of  the  magnetic  field  (circles),
    together with  a best  fit by a  damped oscillation  $A\cos \omega
    te^{\gamma  t}$ (solid  line). The  values of  the parameter  $A$,
    $\gamma$ and $\omega$ are reported in  each panel. The fit is very
    good either below  and above $\beta=2$. Notice  that $A$, $\gamma$
    are essentially  independent from  $\beta$, while  $\omega$ agrees
    well     with     the      cyclotron     frequency     $\omega_c$.
  }\label{fig:corr_vel}
\end{figure}

The law $D_\perp = \frac {\gamma A}{\gamma^2 + \beta^{2}}$, was not
obtained by interpolation, but by the use of the following argument.
Let us introduce the velocity auto correlation $\langle \vec
{v}_{\perp}(t) \cdot \vec{v}_{\perp} (0)\rangle$ where
$\vec{v}_{\perp}$ in the component of the mean particles velocity
transverse to the magnetic field, and the brackets denote the
time averages along the orbit. Then the diffusion coefficient can be
expressed (see the handbook~\cite{wannier_statistical_2010}) as
\begin{equation}\label{eq:diffusione}
  D_{\perp} = \frac 12\int_0^{+\infty} \langle \vec {v}_{\perp}(t)
  \cdot \vec{v}_{\perp} (0)\rangle \diff t
\end{equation}
whenever the velocity auto correlation decays at $t\to +\infty$ fast
enough. Let us introduce a function $f(t)$ to describe this decay
by setting
\begin{equation}\label{eq:vel_corr}
\langle \vec {v}_{\perp}(t) \cdot \vec{v}_{\perp} (0)\rangle \simeq
\langle \vec {v}_{\perp}(0) \cdot \vec{v}_{\perp} (0)\rangle
\cos(\omega_ct) f(t) \ .
\end{equation}

\begin{figure}[t]
  \centering
  \includegraphics[width=\textwidth]{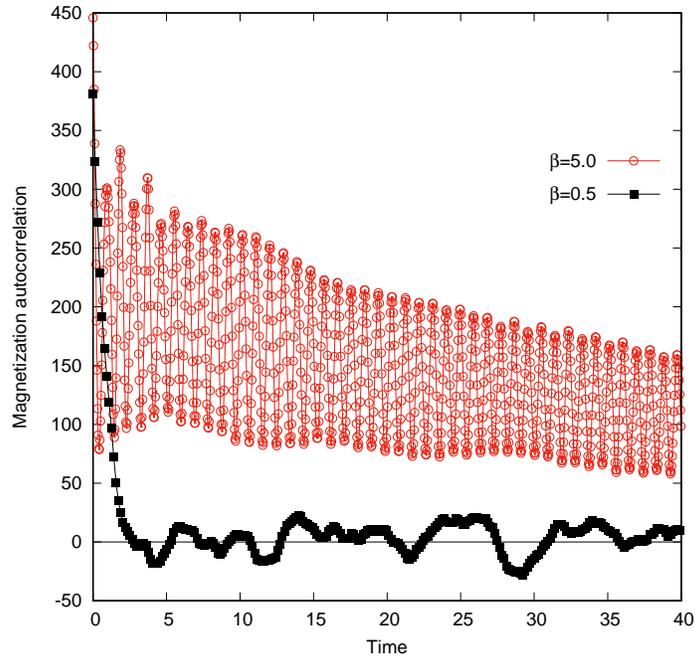}
  \caption{ Autocorrelation of the magnetization (along the magnetic
    field direction) as a function of time, for two different values
    $\beta=0.5$ (square) and $\beta=5$ (circles) of the magnetic
    field. Notice that, for a field below the threshold the
    auto correlation vanishes very fast, while, above the threshold,
    autocorrelation vanishes eventually on a totally different time
    scale.}\label{fig:magn}
\end{figure}
\begin{figure}
  \centering
  \includegraphics[width=\textwidth]{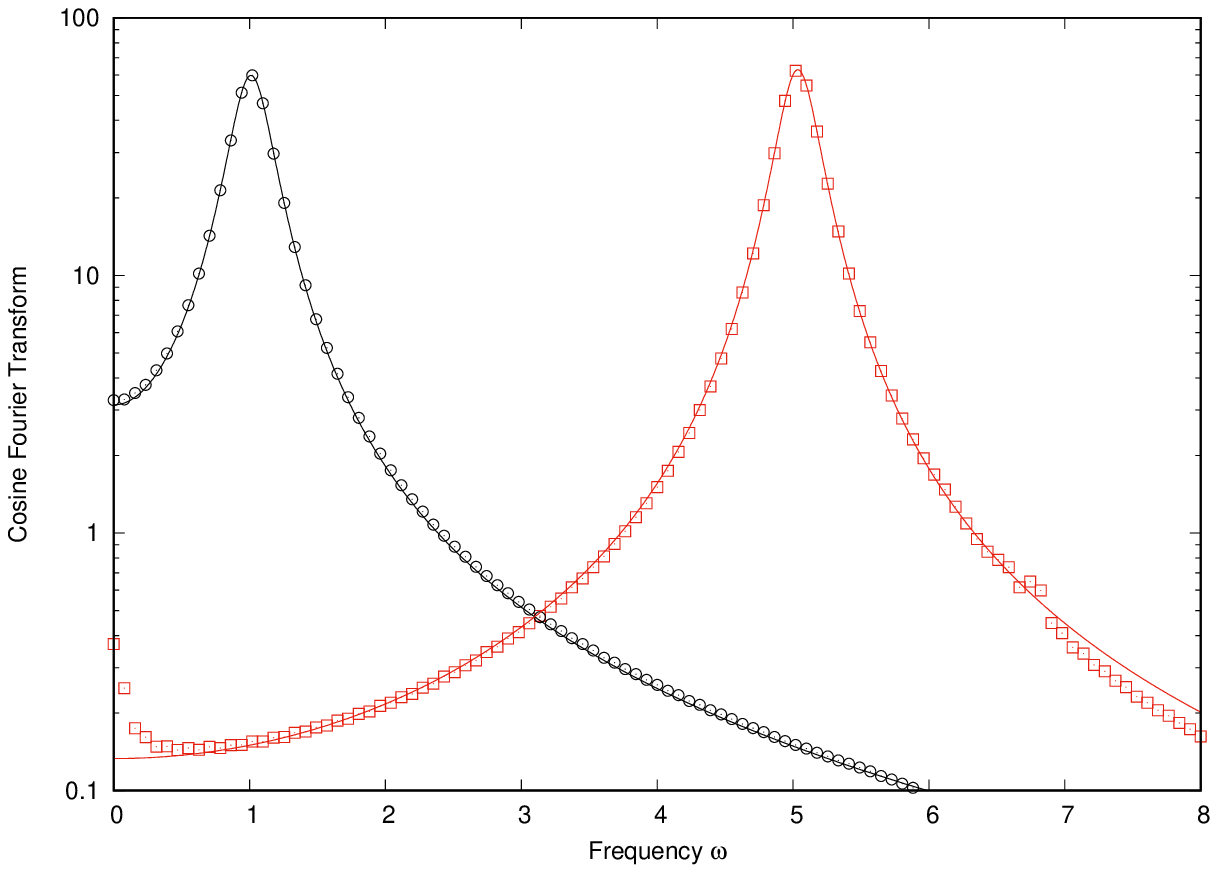}
  \caption{ \label{fig:spettri} Spectrum of the velocity as a function
    of $\omega$ in semi logarithmic scale, for two different value of
    $\beta=0.5$ (circle) and $\beta=5$ (square). The continuous lines
    are the Fourier transform of the damped oscillation
    $2Ae^{-\gamma t}\cos\omega t $. For $\beta=1.0$ the slope of the spectrum
    vanishes at $\omega=0$; for $\beta=5$ the slope remains different
    from zero.  This latter case hints to a decay of the velocity
    auto correlation as $1/t^2$ for $t\to+\infty$ (see text). }
\end{figure}
Recalling the electronic dynamics, we expect that, due that the
gyration along the magnetic field lines alone,
$\langle \vec {v}_{\perp}(t) \cdot \vec{v}_{\perp} (0)\rangle$
would oscillate with the cyclotron frequency $\omega_c$.
However, the electric
interaction between the electrons instead determines a loss of 
coherence of the electronic motion, and thus the decaying to
zero of the auto correlation. A common choice is to consider an
exponential decay, and thus \mbox{$f(t)=2Ae^{-\gamma t}$},
where $\gamma$ is a parameter representing the inverse
of the decorrelation time. Then formula \eqref{eq:diffusione}
gives exactly $D_\perp = \frac{\gamma A}{\gamma^2 + \beta^{2}}$.

In figure~\ref{fig:corr_vel} the velocity auto correlation is
plotted as a function of time, for different values of
$\beta$, either above and below the threshold $\beta=2$. As one can
check the law (\ref{eq:vel_corr}) is nicely agreed. From the values
reported in table~\ref{table:uno}, one finds that the parameters $A$
and $\gamma$ are quite independent from $\beta$, while $\omega$ turns
out to be very close to the cyclotron frequency $\omega_c$ (equal to
$\beta$ in our units) as expected. So, taking average values $A=9.9$
and $\gamma=0.168$, and $\omega=\omega_c$ the values of the diffusion
coefficient $D_\perp$ obtained by numerical computations can be
recovered from formula (\ref{eq:diffusione}).

Now, a peculiar fact happens. In fact, while formula
(\ref{eq:vel_corr}) appears to be in very good agreement with the
velocity auto correlation plots for all the values of $\beta$
considered, the formula $D_\perp = \frac {\gamma A}{\gamma^2 +
  \beta^{2}}$ is valid only below a threshold
$\beta_{cr}\simeq2$. This notwithstanding, if we compute the time
integral numerically and inject it into formula~(\ref{eq:diffusione}),
the resulting values of $D_{\perp}$ agree very well with those found
using linear regression, as one can check from the
Table~\ref{table:uno} comparing the values reported in the second and
third column. Notice that, when computing the integral appearing
formula~(\ref{eq:diffusione}), one has to truncate the integral to an
upper limit $t_{max}$ chosen carefully, i.e., not to large, otherwise
the numerical errors introduced in computing the auto correlation $
\langle \vec {v}_{\perp}(t) \cdot \vec{v}_{\perp} (0)\rangle$ spoil
the result.

A possible explanation of this transition when passing from a weakly
magnetized to a strongly magnetized regime can be given in the spirit
of the paper~\cite{carati_transition_2012}, were it was surmised
that at low $\beta$ the plasma is in a fully chaotic regime, but as
$\beta$ is increased a transition to a partially ordered regime
occurs.

It was shown in that paper how, in such a partially ordered regime, a
perturbation theory could apply by adapting the Hamiltonian
perturbation theory developed for system of interest in statistical
mechanisc (see for example Refs.~\cite{C2007, CM2012,
  GPP2015,huveneers}) to the case of plasma. The idea is that, in the
thermodynamical limit, one cannot controll the adiabatic invariants
along each individual trajectory (in the phase space), but it is
instead possible to controll their auto correlations with respect to
an invariant measure, showing that they vanish exponentially slow in
the perturbative parameter.  Notice that, in virtue of the linear
response theory, such auto correlations correspond often to important
physical observables.

In the above mentioned paper~\cite{carati_transition_2012}, it
was considered the case of the component $M$ of the magnetization
along the magnetic field $\vec{B}$, defined (as usual) as
\begin{equation}
  M \dfnd \frac e{2mc}  \frac 1N\sum_k (\vec{v}_k \times \vec{x}_k)_z \ .
\end{equation}
Notice that, the auto
correlation $\mathcal{C}_M(t)\dfnd <M(t)M(0)>$ is an important
quantity because, according the linear response theory, its
Fourier transform $\hat{\mathcal{C}}_M(\omega)$ gives the magnetic
susceptibility $\chi(\omega)$ at the frequency $\omega$ (see for
example Refs.~\cite{kubo,klimontovic} or the Appendix B of
Ref.~\cite{BCG2011}).

In our case, the behavior the auto correlation is different below and
above the threshold.  This is shown in Figure~\ref{fig:magn}: for low
magnetic field the auto correlation relaxes to zero, while for high
magnetic field it keeps oscillating and eventually vanishes on a
totally different time scale. So the magnetizazion could be considered
an adiabatic invariant of the system, thus implying that the dynamics
remains partially correlated for long times.

A similar mechanism could be at work also in the case of the velocity
auto correlation, even if from our plots this is not as evident as in
the case of the magnetization. In fact, a clue can be obtained by
looking at the spectrum of the velocity auto correlation, i.e., at
their cosine Fourier transform, as shown in figure~\ref{fig:spettri}
(in semi logarithmic scale). Notice that $D_{\perp}$ is simply half the
value of the spectrum at $\omega=0$.

For $\beta=0.5$ the slope of the spectrum seems to vanish at
$\omega=0$. As the spectrum is an even function of the frequency this
is coherent with the behavior of a smooth function. Instead, for
$\beta=5$, the spectrum slope seems to remain (negative and) different
from zero at $\omega=0$. Now, it can be very easily shown that, by
denoting with $\hat f(\omega)$ the Fourier cosine transform of the
function $f(t)$ then, for $t\to+\infty$, one has $f(t) = \frac 2\pi
\frac 1{t^2}\left( -\hat f'(0) + o(1) \right )$, if all the
derivatives of $\hat f(\omega)$ up to the second one are
integrable.\footnote{ This can be inferred by the inverse transform
  formula $f(t)=\frac 2\pi \int_0^{+\infty} \hat f(\omega)\cos \omega
  t \diff \omega $ by integrating two times by part and using the
  Riemann--Lebesgue lemma on the remainder.} So it seems that in our
data on the velocity auto correlation, there is a small component
which decays very slowly (i.e., as $t^{-2}$) to zero.  Nevertheless
such a component gives a very big contribution to the diffusion
coefficient (more then the double of the one due to the exponentially
decreasing part $2Ae^{-\gamma t}\cos \omega t $).

In any case, at the moment, it is not clear, for what reason, in a
less chaotic regime, the diffusion coefficient apparently decreases,
as a function of $\beta$, at a slower rate with respect to the fully
chaotic regime.

\section{\label{sec:conclusions}Final considerations}

In this work we performed MD simulations of a magnetized electron
gas, also called a One Component Plasma.  We have shown that
such a system shows a transition between two different regimes
as the value of the parameter $\beta$ is raised above a
critical threshold of about $2$.

The transition occurs both on a macroscopic level, with a change in
the diffusive behavior, and on a microscopic level as well: when the
parameter is raised, the system passes from a chaotic state to a
partially ordered one. This is the main finding that we have pointed
out, and is a rather new phenomenon, only remotely addressed in the
literature, up to now. As a matter of fact a similar result, for what
concerns the behavior of the diffusion coefficient, was published
quite recently in Ref.~\cite{vidal_extended_2021}. But the
authors tried to explain this phenomenon in the framework of the
kinetic theory, looking at the behavior of the particles during the
``collisions''.  We refrain from this approach, and we try to discuss
it according to the ergodic theory of dynamical system, using its
standard tools. In particular, our aim is to understand if the
dynamics is truly chaotic or not, and, in this latter case, what are
the consequence for what concerns the macroscopic quantity
characterizing the plasma.

As regards the direct consequences of our results on plasma
physics, a strong objection may be raised on the dependence, we have
not explored, of the results on the number of particles $N$. In
particular, in Ref.~\cite{vidal_extended_2021} it was
claimed that a few hundred particles are sufficient in the low $\beta$
regime, but for high $\beta$ a huge number of particles (above $10^6$)
is needed. In particular, they show that the value of the diffusion
coefficient slowly decreases as $N$ is increased. But it seems
unlikely that those values would ever agree with the kinetic law
\mbox{$D_\perp \propto \beta^{-2}$}, although a Bohm relation
\mbox{$D_\perp\propto\beta^{-1}$} may finally show up in the high
$\beta$ regime, as in Ref.~\cite{ott_diffusion_2011}.  Also,
such a high value of $N$ is going towards the real number of
particles, so at this point even the use of periodic boundary
conditions become questionable. Finally, maybe the problem is that the
diffusive regime is not normal, so that the diffusion coefficient is
not well defined. This problem is discussed at length in
Ref.~\cite{balescu_aspects_2005}.

In any case, we think that there are much more fundamental questions
to address about the portability of our results to real plasmas:
above all, the absence of positive ions in our models. We hope to be
able to address such a big issue in the (near) future.

\section*{Data Availability Statement}

The data that support the findings of this study are available from the
corresponding author upon reasonable request


\bibliography{diffusionPlasma_mod}
\bibliographystyle{plain}




\end{document}